# NaZnF$_3$ as a low-pressure analogue of MgSiO$_3$

Dominik Kurzydłowski,[1*] Arkadiusz Gajek,[2] Zoran Mazej,[3]

[1] Faculty of Mathematics and Natural Sciences, Cardinal Stefan Wyszyński University in Warsaw, 01-038 Warsaw, Poland;

[2] Institute of Physical Chemistry, Polish Academy of Sciences, 01-224 Warsaw, Poland;

[3] Department of Inorganic Chemistry and Technology, Jožef Stefan Institute, SI-1000 Ljubljana, Slovenia

* d.kurzydlowski@uksw.edu.pl

Solid-state systems whose properties at high pressure (exceeding 1 GPa) mimic those of MgSiO$_3$ are of large importance in the study of the interior of planets. By means of Density Functional Theory (DFT) calculations we studied the high-pressure properties of a MgSiO$_3$ analogue, NaZnF$_3$. We reproduce the phase-transition sequence previously reported for this compound (GdFeO$_3$ → CaIrO$_3$ → La$_2$S$_3$), and predict that it should undergo a two-step dissociation: decomposition into a equimolar mixture of Na$_2$ZnF$_4$ and NaZn$_2$F$_5$ at 25.4 GPa, followed by a breakdown into ZnF$_2$ and NaF at 66.8 GPa. These processes are analogous to those predicted for compressed MgSiO$_3$. Moreover, both Na$_2$ZnF$_4$ and NaZn$_2$F$_5$ are isostructural with analogous phases from the Mg-Si-O system. We also find that both these novel compounds are thermodynamically stable at ambient conditions (Na$_2$ZnF$_4$) or at low pressure of 19 GPa (NaZn$_2$F$_5$). Our study indicates that NaZnF$_3$ could serve as a good low-pressure analogue of MgSiO$_3$ exhibiting the same sequence of phase transitions, and pressure induced decomposition, but at pressures an order of magnitude lower.

## I. INTRODUCTION

The perovskite (GdFeO$_3$-type) and post-perovskite (CaIrO$_3$-type) polymorphs of MgSiO$_3$ are the major constituents of the Earth's mantle. It is assumed that they are also present in the interior of other rocky planets, including super-Earths. Therefore, modelling the properties of the interior of these objects requires knowledge on the properties of silicates at conditions found there (pressures exceeding 500 GPa) [1]. This knowledge is acquired mainly through laboratory experiments, in most cases utilizing the diamond anvil cell (DAC). The importance of high-pressure experiments is exemplified by the discovery of the perovskite to post-perovskite transition in MgSiO$_3$ at 125 GPa, which explained the unusual properties of the lowermost part of the Earth's mantle (the D″ layer) [2–4].

Despite the impressive progress in experimental techniques [5–10], recreating the extreme conditions of the interiors of planets in the laboratory is still very challenging. For that reason *ab initio* modelling [11–14], as well as experiments on MgSiO$_3$ analogs, are often used in the study of the deep Earth [15]. The so-called low-pressure analogs (LPAs) of MgSiO$_3$ are systems which display similar chemistry, phase transition sequence, and structure-property relations, but at lower pressures (typically below 50 GPa). With the increasing number of newly discovered exoplanets [16], the study on LPAs has gained much attention with several systems proposed as MgSiO$_3$ analogs, such as NaMgF$_3$ [17–25], KZnF$_3$ [26–29], and MgGeO$_3$ [30–33]. The interest in the high-pressure behavior of LPAs was mostly focused on the pressure-induced phase transitions in the parent $ABX_3$ system [34]. However, recent theoretical studies indicated that at pressures exceeding 750 GPa (0.75 TPa) the post-perovskite phase of $MgSiO_3$ should decompose via a three-step process, with the formation of $Mg_2SiO_4$ and $MgSi_2O_5$ in the first step [35–37]. This shifted the focus towards pressure-induced decomposition of $MgSiO_3$ analogues [24,25].

Here we present a computational investigation, based on solid state Density Functional Theory (DFT), on the ability of $NaZnF_3$ to exhibit the same phase transition sequence and pressure-induce decomposition as $MgSiO_3$, but at much lower pressures. We also analyze the geometry of the thermodynamically stable compositions of this compound at high pressure, and make a comparison between $NaZnF_3$ and two other recently proposed LPAs, $NaMgF_3$ and $MgGeO_3$.

## II. COMPUTATIONAL METHODS

Periodic DFT calculations of the geometry and enthalpy of high-pressure polymorphs of $NaF$, $ZnF_2$, $NaZnF_3$, $Na_2ZnF_4$, and $NaZn_2F_5$ utilized the PBEsol functional [38], as implemented in CASTEP (version 19.11) [39]. We found that for $NaZnF_3$ the chosen method yields phase transition pressures, and the pressure dependence of the unit cell vectors in line with experiment – see Fig. S1 and Table S1 in the Supplemental Material [40]. Good agreement between theory and calculations was also found for the Raman spectrum of the ambient-pressure $GdFeO_3$ structure of $NaZnF_3$ (Fig. S2) [40].

The valence electrons were described with a plane-wave basis set (1100 eV cut-off), while norm-conserving pseudopotentials were used for the description of core electrons (Na: $2s^2 2p^6 3s^1$, F: $2s^2 2p^5$, Zn: $3d^{10} 4s^2$). The convergence criterion for the electronic minimization was $10^{-7}$ eV per atom. Sampling of the Brillouin zone was done through a Monkhorst–Pack mesh [41], with a $2\pi \times 0.03$ Å$^{-1}$ spacing of $k$-points. Geometry optimization of the crystal structures was performed with the use of the Broyden–Fletcher–Goldfarb–Shanno scheme [42]. Structures were optimized until the following convergence criteria were met: (i) forces acting on the atoms were smaller than 5 meV/Å; (ii) difference between the applied hydrostatic pressure and all stress components was smaller than 0.05 GPa; (iii) the maximum ionic displacement smaller than $5 \cdot 10^{-4}$ Å.

Evolutionary algorithm searches were performed to identify the lowest-enthalpy structures of $NaZnF_3$, $Na_2ZnF_4$, and $NaZn_2F_5$ at high pressure (structural data for the relevant phases is given in Table S2 [40]). Searches were performed with the use of the XtalOpt software (version r12) [43], coupled with DFT calculations utilizing the PBEsol functional (conducted in the VASP software) [44,45]. Searches were done at 10, 40, 80 GPa, and 120 GPa for $Z$ up to 4. For $NaZn_2F_5$ an additional search at 10 GPa was conducted for $Z$ up to 8.

We also used CASTEP for calculating the phonon dispersion curves (with a $2\pi \times 0.05$ Å$^{-1}$ q-point spacing) using density-functional perturbation theory (DFPT) [46]. We used a fine FFT grid (CASTEP keywords: grid_scale : 2.5, fine_grid_scale : 3.5) and more restrictive SCF convergence criterion ($2 \cdot 10^{-10}$ eV). Phonon dispersion curves for the relevant phases at selected pressures are given in Fig. S3 [40].

Thermodynamic stability of various polymorphs of NaF, NaZnF$_3$, Na$_2$ZnF$_4$, and NaZn$_2$F$_5$ was judged by comparing their enthalpy ($H$), and thus the calculated phase transitions pressures formally correspond to $T = 0$ K at which the Gibbs free energy ($G = H - S \cdot T$, where $S$ is the entropy) is equal to the enthalpy. Phonon dispersion calculations, performed at selected pressures, confirmed the dynamic stability of the studied phases within their thermodynamic stability window, and enabled calculation of the zero-point energy (ZPE) contribution. The inclusion of ZPE does not influence markedly the phase stability of the studied compounds (see Section III). Visualization of all structures was performed with the VESTA software package [47]. For symmetry recognition we used the FINDSYM program [48].

## III. RESULTS AND DISCUSSION

Before we move to the phase transitions and thermodynamic stability of ternary phases in the Na-Zn-F system we address the computational results for the binary fluorides: NaF, ZnF$_2$. For the former we predict a phase transition from the NaCl-type (B1) structure to the CsCl-type (B2) polymorph at 23.5 GPa – in close accordance with the experimental value of 27 GPa [49]. At ambient conditions ZnF$_2$ adopts the TiO$_2$ (rutile) structure. Upon compression the following sequence of phase transitions is predicted by our calculations: TiO$_2 \rightarrow$ CaCl$_2 \rightarrow$ HP-PdF$_2 \rightarrow$ HP1-AgF$_2 \rightarrow$ PbCl$_2$ (cotunnite), with the Zn$^{2+}$ coordination increasing from 6 (TiO$_2$, CaCl$_2$, HP-PdF$_2$) through 7 (HP1-AgF$_2$) to 9 (PbCl$_2$) [50–52]. These results are in good agreement with recent experimental and computational studies on ZnF$_2$ [52,53].

### A. NaZnF$_3$

Our calculations also reproduce previous experimental and theoretical results on the high-pressure phase transitions of NaZnF$_3$ [54–56]. At ambient conditions (effectively 0 GPa) we find the distorted perovskite GdFeO$_3$-type structure (*Pnma* symmetry) as the most stable polymorph of NaZnF$_3$ (Fig. 1) – in line with experiment. This structure exhibits octahedral coordination of Zn$^{2+}$ by F$^-$ anions; the coordination number (CN) of Na$^+$ is 8. The GdFeO$_3$ structure is predicted to transform to a post-perovskite (CaIrO$_3$-type, *Cmcm* symmetry) polymorph at 7.7 GPa, in close accordance to the experimental value of 5 GPa (at room temperature) [55]. The CaIrO$_3$ structure exhibits the same CNs of the cations as GdFeO₃, but differs in the connectivity of their coordination polyhedral. In GdFeO₃ each $ZnF_6$ octahedra shares all of its corners with one neighbor thus forming a 3D network, while in CaIrO₃ corner and edge sharing leads to a layered network [Fig. 2(a)]. We find that CaIrO₃ is dynamically stable even at 1 atm [40] – in accordance with the observed metastability of this polymorph upon pressure quenching [55].

At 23.4 GPa the CaIrO₃ phase is predicted to transform to a La₂S₃-type structure (*Pnma* symmetry). This structure, sometimes referred to as Gd₂S₃-type, was recently proposed as a post-post-perovskite phase in many ABO₃, ABF₃, and A₂O₃ compounds [34], including NaZnF₃ [56]. At 23.4 GPa the La₂S₃ polymorph exhibits CN(Zn) and CN(Na) equal to 6 and 8, respectively. However, the coordination environment of $Zn^{2+}$ is distorted from the octahedron [Fig. 2(b)]. Pairs of these distorted octahedrons are arranged into 1D chains running along the *b* axis. Upon compression an additional $F^−$ anion enters the coordination sphere of $Zn^{2+}$, resulting in a change of the coordination polyhedron from a distorted octahedron to a distorted mono-capped trigonal prism (CN = 7). At the same time the coordination number of $Na^+$ increases to 9. This pressure-induced evolution of the coordination sphere of $Zn^{2+}$ mimics that occurring in the HP1-AgF₂-type high-pressure polymorph of ZnF₂ [52]. The seven-fold coordination of $Zn^{2+}$ persists in the La₂S₃ structure up to 100 GPa, and this polymorph remains the ground-state structure of NaZnF₃ up to that pressure.

High-pressure experiments indicated that above 25 GPa the CaIrO₃ polymorph of NaZnF₃ transforms reversibly into a novel, unidentified phase [54]. Based on DFT calculations Cheng *et al.* proposed that this new phase should be isostructural to Sb₂S₃ (sometimes referred to as the U₂S₃-type structure, *Pnma* space group) which they call *pPv-Pnma* [56]. However, they find that this structure has a higher enthalpy than the La₂S₃ phase (termed by them *ppPv-Pnma*) in the relevant pressure range. As can be seen in Fig. 1 our calculations confirm this finding.

Just above the CaIrO₃ – La₂S₃ phase transition, at 25.4 GPa, NaZnF₃ is predicted to become thermodynamically unstable with respect to an equimolar mixture of Na₂ZnF₄ and NaZn₂F₅ (Fig. 1):

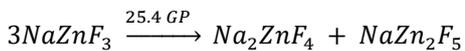

$$3NaZnF_3 \xrightarrow{25.4\ GP} Na_2ZnF_4 + NaZn_2F_5$$

One may view Na₂ZnF₄ as NaF-rich, and NaZn₂F₅ as ZnF₂-rich with respect to NaZnF₃. An analogous decomposition path (yielding Mg₂SiO₄ and MgSi₂O₅) is predicted to occur in MgSiO₃ at 750 GPa [37]. As we will show below the decomposition products of NaZnF₃ are predicted to be isostructural with

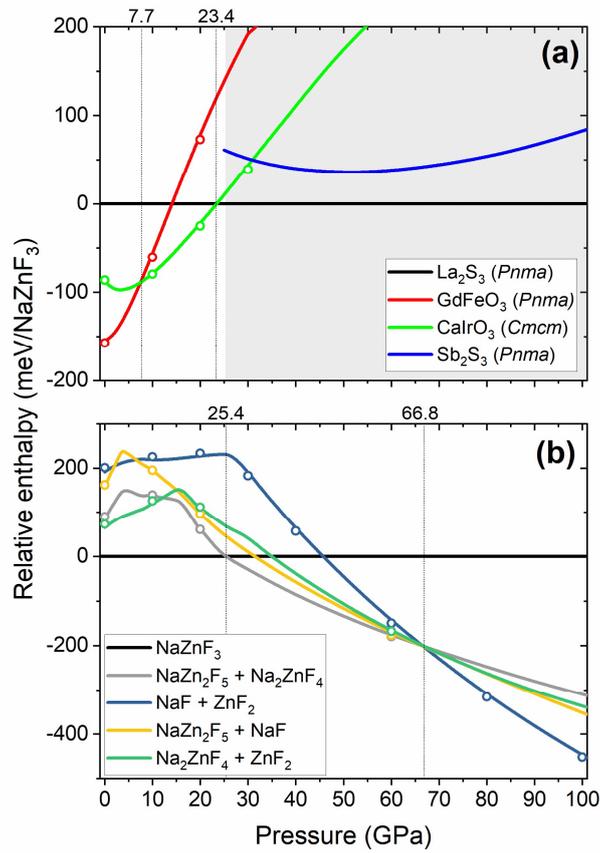

**FIG. 1** (a) Pressure dependence of the relative enthalpy of NaZnF₃ polymorphs with respect to the enthalpy of the La₂S₃ structure (grey area depicts the region in which NaZnF₃ is thermodynamically unstable); (b) relative enthalpy of NaZnF₃ dissociation products (within their most stable phases) with respect to NaZnF₃. In both (a) and (b) dots indicate ZPE-corrected values. Vertical lines in (a) indicate phase transition of NaZnF₃ (at 7.7 and 23.4 GPa); vertical lines in (b) indicate the pressure of decomposition of NaZnF₃ into Na₂ZnF₄ and NaZn₂F₅ (25.4 GPa), and the subsequent decomposition of these two compounds into NaF and ZnF₂ (at 66.8 GPa)

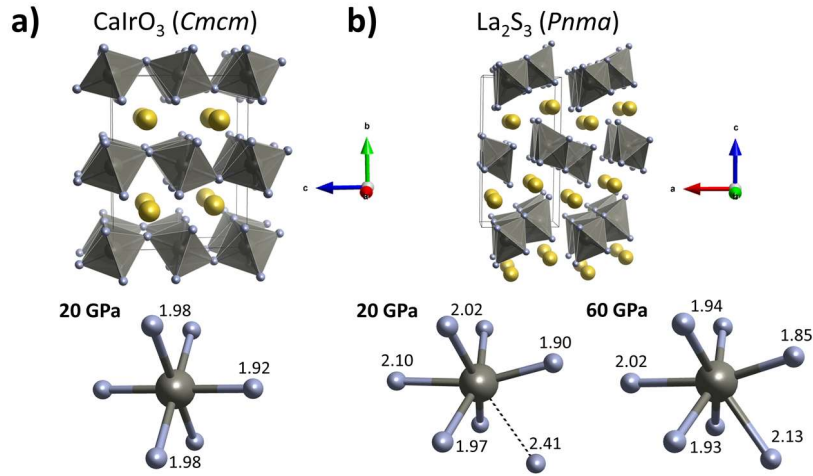

**FIG. 2** The (a) CaIrO₃ and (b) La₂S₃ polymorphs of NaZnF₃. Yellow/grey/blue balls denote Na/Zn/F atoms. Atomic distances within the Zn polyhedron at selected pressures are given in Å.

the analogous products of MgSiO$_3$ decomposition. Upon compression another decomposition event is predicted to occur at 66.8 GPa with both Na$_2$ZnF$_4$ and NaZn$_2$F$_5$ fragmenting into binary fluorides:

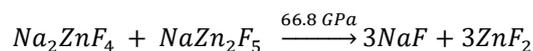

$$Na_2ZnF_4 + NaZn_2F_5 \xrightarrow{66.8\ GPa} 3NaF + 3ZnF_2$$

At this pressure NaF is predicted to adopt the CsCl-type structure, while ZnF$_2$ the PbCl$_2$-type structure.

## B. Na$_2$ZnF$_4$

Our calculations indicate that Na$_2$ZnF$_4$, which although hypothesized has not been reported up to date [57], should be thermodynamically stable even at ambient conditions (Fig. 3). At pressures below 15.6 GPa it is predicted to adopt the Sr$_2$PbO$_4$-type structure of *Pnma* symmetry which consists of chains built from edge-sharing ZnF$_6$ octahedra separated by Na$^+$ cations [Fig. 4(a)]. According to the ICSD database [58], the Sr$_2$PbO$_4$ structure type, together with its lower-symmetry variant (Na$_2$CuF$_4$-type of *P2$_1$/c* symmetry), is adopted at ambient conditions by a range of ternary oxides, but only seven ternary halogens (among them: Na$_2$MgCl$_4$ [59], β-K$_2$AgF$_4$ [60], and Na$_2$CuF$_4$ [61]). It can be viewed as a post-perovskite phase of the layered perovskite K$_2$NiF$_4$-type structure (*n* = 1 member of the Ruddlesden-Popper series) [62]. At 1 atm the energy of the Sr$_2$PbO$_4$ structure is about 260 meV per Na$_2$ZnF$_4$ (= 25 kJ/mol) lower than that of the spinel structure (*Fd$\bar{3}$m*, *Z* = 8) featuring tetrahedrally coordinated Zn$^{2+}$ cations, in accordance with the preference of Zn$^{2+}$ to adopt octahedral coordination at ambient pressure.

However, the Sr$_2$PbO$_4$ structure is predicted to transform at 15.6 GPa into a structure of *I$\bar{4}$2d* symmetry which can be viewed as a distorted variant of the CdMn$_2$O$_4$ spinel [63]. This phase change is connected with a reduction of the CN of Zn$^{2+}$ from 8 to 4 [Fig. 4(b)], while the CN of Na$^+$ increases from 7 to 8. In the *I$\bar{4}$2d* structure each Zn$^{2+}$ cation is surrounded by four F$^-$ anions forming a distorted tetrahedron (Zn-F distances of 1.96 Å at 15 GPa). Next-nearest-neighbor contacts, also forming a tetrahedron around Zn$^{2+}$, are more than 20 % longer. At 15 GPa the *I$\bar{4}$2d* structure is analogous to the predicted high-pressure phase of Ag$_3$F$_4$ (=Ag$^{(I)}_2$Ag$^{(II)}$F$_4$) [64].

The 4-fold coordination evolves into an 8-fold (4+4) one with pressure, as compression induces a substantial shortening of the next-nearest-neighbor Zn-F contacts – at 100 GPa their length is only 10 % larger than that of the nearest-neighbor contacts [Fig 4(b)]. This increase in the CN of Zn$^{2+}$ makes this phase analogous to the *I$\bar{4}$2d* structure of Mg$_2$SiO$_4$ [37,65,66], where a similar 4+4 coordination (with a similar difference in Mg-O lengths) is found at 1 TPa [37]. For a more detailed comparison of the Na$_2$ZnF$_4$ and Mg$_2$SiO$_4$ structures see the Supplemental Material [40].

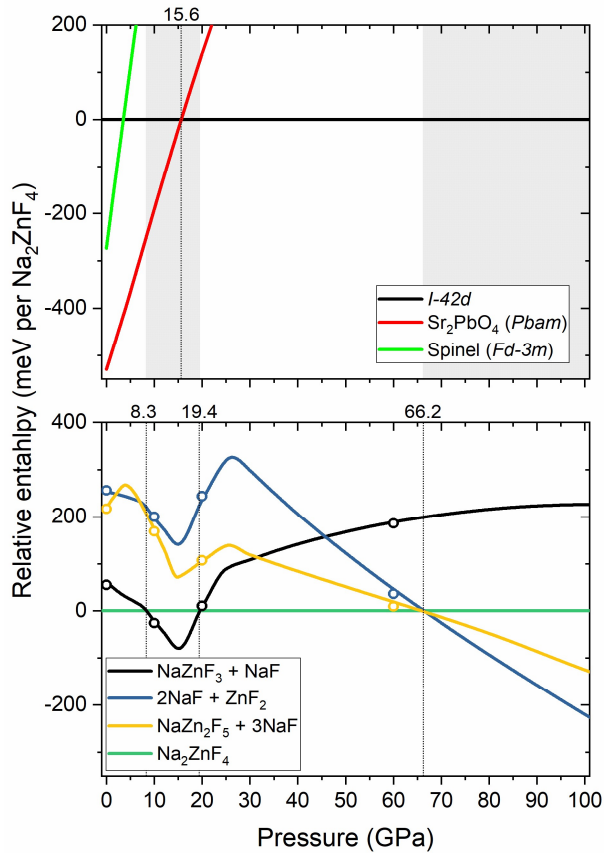

**FIG. 3** (a) Pressure dependence of the relative enthalpy of $Na_2ZnF_4$ polymorphs with respect to the enthalpy of the $I\bar{4}2d$ structure (grey area depicts the region in which $Na_2ZnF_4$ is thermodynamically unstable); (b) relative enthalpy of $Na_2ZnF_4$ dissociation products (within their most stable phases) with respect to $Na_2ZnF_4$. In both (a) and (b) dots indicate ZPE-corrected values. The vertical line in (a) indicates the phase transition in $Na_2ZnF_4$ at 15.6 GPa; vertical lines in (b) indicate the pressure of decomposition of $Na_2ZnF_4$ into $NaZnF_3$ and NaF (8.3 GPa), it's subsequent re-emergence from this mixture (19.4 GPa) and decomposition into NaF and $ZnF_2$ (at 66.2 GPa).

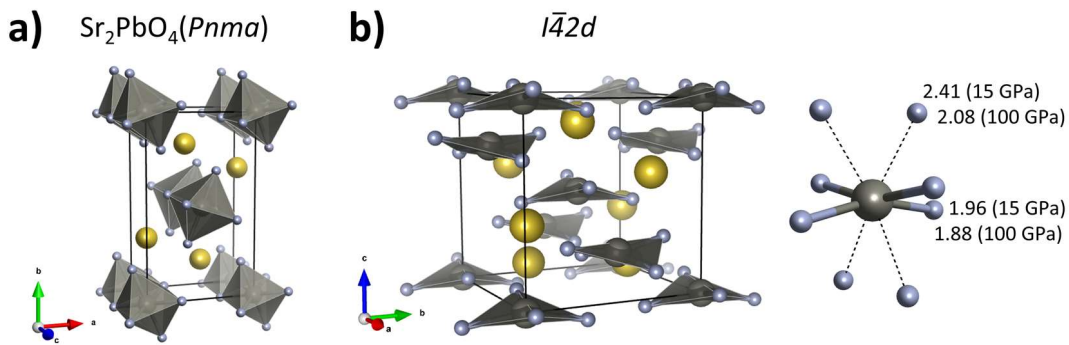

**FIG. 4** The (a) $Sr_2PbO_4$ and (b) $I\bar{4}2d$ polymorphs of $Na_2ZnF_4$. Nearest-neighbor and next-nearest-neighbor Zn-F distances in the $I\bar{4}2d$ structure at 15 and 100 GPa are given in Å. For clarity the unit cell of the $I\bar{4}2d$ polymorph was shifted by $(0,0, \frac{1}{2})$.

Our calculations indicate that at 8.3 GPa $Na_2ZnF_4$ in the $Sr_2PbO_4$ structure should become thermodynamically unstable with respect to decomposition into $NaZnF_3$ ($CaIrO_3$-type) and NaF (CsCl-type).

However, after the $Sr_2PbO_4 \rightarrow I\overline{4}2d$ phase transition $Na_2ZnF_4$ regains its thermodynamic stability at 19.4 GPa. Finally, it is predicted to decompose into binary fluorides (NaF – NaCl, $ZnF_2$ – $PbCl_2$) at 66.2 GPa, in analogy to what was found for the $NaZnF_3$ system:

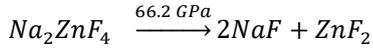

$$Na_2ZnF_4 \xrightarrow{66.2\ GPa} 2NaF + ZnF_2$$

In order to verify experimentally whether $Na_2ZnF_4$ could be obtained at ambient pressure we have performed a solid-state high-temperature reaction between $ZnF_2$ and NaF in a 1:2 mole ratio. This did not lead to formation of $Na_2ZnF_4$ – the obtained product was $NaZnF_3$ in the $GdFeO_3$-type structure accompanied by excess NaF (see Fig. S4 [40]). The discrepancy between this result and the theoretical prediction might lie in temperature effects. However modelling those, for example with the use of the quasi-harmonic approximation [67], lies beyond the scope of the current manuscript.

## C. $NaZn_2F_5$

Calculations indicate that $NaZn_2F_5$ is thermodynamically stable between 18.6 and 66.7 GPa (Fig. 5). Below 18.6 GPa this compound is predicted to decompose into $NaZnF_3$ ($CaIrO_3$) and $ZnF_2$ (HP-$PdF_2$), above 66.7 GPa it should dissociate into binaries (NaF – B2, $ZnF_2$ – $PbCl_2$), as was the case for $NaZnF_3$ and $Na_2ZnF_4$.

In its thermodynamic stability window $NaZnF_3$ should adopt a $P2_1/c$ ($Z = 4$) structure featuring $Zn^{2+}$ cations in two coordination environments: distorted square antiprism (CN = 8) and distorted octahedron (CN = 6), as shown in Fig. 6. Upon compression the coordination number of the latter site increases to 7, in analogy to what was found for the $La_2S_3$ phase of $NaZnF_3$. Both at low and high pressure $Na^+$ is 9-fold coordinated in the $P2_1/c$ structure. This polymorph of $NaZn_2F_5$ is isostructural to the predicted ground state structure of $MgSi_2O_5$ [36].

At low pressure the $P2_1/c$ structure is dynamically stable down to 10 GPa. At this pressure a phonon instability develops at Z point, that is the (½, 0, 0,) wavevector. The distortion resulting from this instability leads to formation of a structure with a doubled unit cell ($Z = 8$) of $Pnma$ symmetry. This polymorph can be related to $P2_1/c$ via rotations of the $Zn^{2+}$ polyhedra around the $b$ cell vector. The $Pnma$ structure is dynamically stable at ambient pressure hinting at the possibility of quenching to 1 atm $NaZn_2F_5$ synthesized at high pressure.

Despite the fact that at 1 atm $NaZn_2F_5$ is unstable against decomposition into $NaZnF_3$ and $ZnF_2$, the enthalpy change associated with the formation of this compound from a 1:2 mixture of NaF and $ZnF_2$ at this pressure is slightly negative (–7.5 kJ/mol). Therefore we have made an attempt to perform the synthesis of $NaZn_2F_5$ from NaF and $ZnF_2$ in a 1:2 mole ratio. However, this resulted in the formation of NaZnF$_3$ and excess ZnF$_2$ (Fig. S4 [40]), in accordance with the predicted thermodynamics. It seems that a high-pressure synthesis (p > 20 GPa) is required for obtaining NaZn$_2$F$_5$.

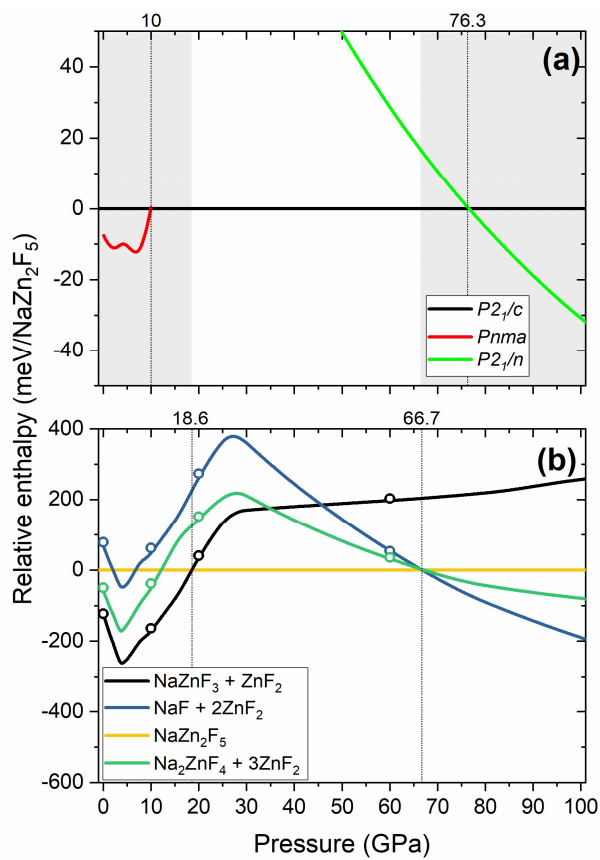

FIG. 5 (a) Pressure dependence of the relative enthalpy of NaZn$_2$F$_5$ polymorphs with respect to the enthalpy of the *P2$_1$/c* structure (grey area depicts the region in which NaZn$_2$F$_5$ is thermodynamically unstable); (b) relative enthalpy of NaZn$_2$F$_5$ dissociation products (within their most stable phases) with respect to NaZn$_2$F$_5$. In (b) dots indicate ZPE-corrected values. The vertical lines in (a) indicates the phase transition in NaZn$_2$F$_5$ at 10 and 76.3 GPa; vertical lines in (b) indicate the pressure of formation of Na$_2$ZnF$_4$ from NaZnF$_3$ and ZnF$_2$ (18.6 GPa), and its decomposition into NaF and ZnF$_2$ (at 66.7 GPa)

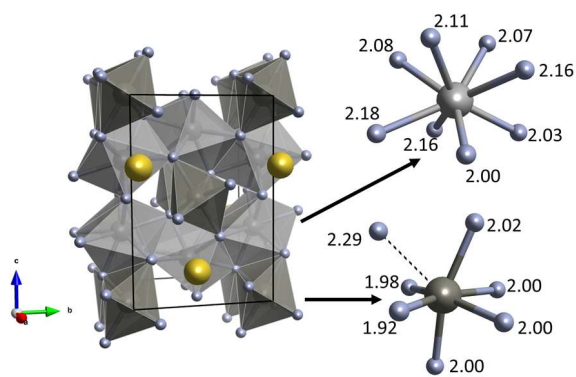

FIG. 6 The *P2$_1$/c* structure of NaZn$_2$F$_5$. Different shades of grey indicate different Zn sites; Zn-F distances at 20 GPa are given in Å.

To summarize the presented results below we give the summary of the ground-state phase transition in the NaZnF₃, Na₂ZnF₄, and NaZn₂F₅ systems. Phases that up to date have not been obtained experimentally are given in bolded font.

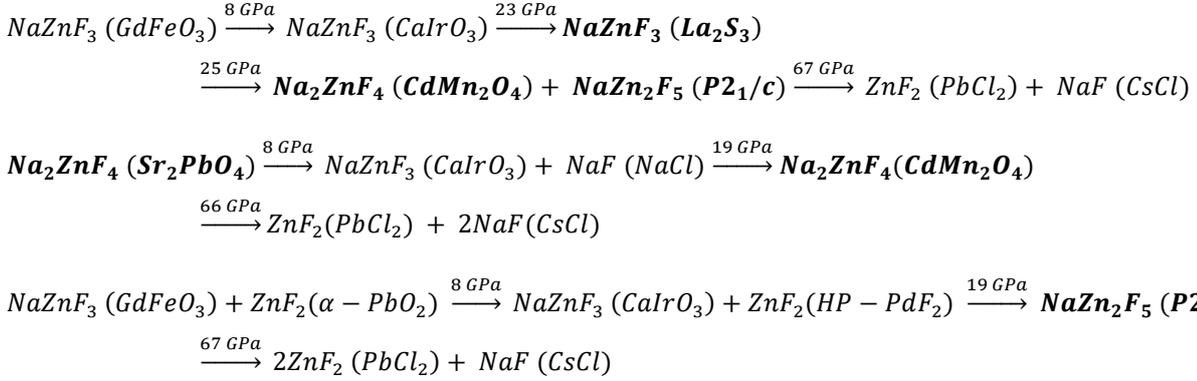

$$NaZnF_3\ (GdFeO_3) \xrightarrow{8\ GPa} NaZnF_3\ (CaIrO_3) \xrightarrow{23\ GPa} \boldsymbol{NaZnF_3\ (La_2S_3)}$$
$$\xrightarrow{25\ GPa} \boldsymbol{Na_2ZnF_4\ (CdMn_2O_4)} + \boldsymbol{NaZn_2F_5\ (P2_1/c)} \xrightarrow{67\ GPa} ZnF_2\ (PbCl_2) + NaF\ (CsCl)$$

$$\boldsymbol{Na_2ZnF_4\ (Sr_2PbO_4)} \xrightarrow{8\ GPa} NaZnF_3\ (CaIrO_3) + NaF\ (NaCl) \xrightarrow{19\ GPa} \boldsymbol{Na_2ZnF_4(CdMn_2O_4)}$$
$$\xrightarrow{66\ GPa} ZnF_2(PbCl_2)\ +\ 2NaF(CsCl)$$

$$NaZnF_3\ (GdFeO_3) + ZnF_2(\alpha - PbO_2) \xrightarrow{8\ GPa} NaZnF_3\ (CaIrO_3) + ZnF_2(HP-PdF_2) \xrightarrow{19\ GPa} \boldsymbol{NaZn_2F_5\ (P2_1/c)}$$
$$\xrightarrow{67\ GPa} 2ZnF_2\ (PbCl_2) + NaF\ (CsCl)$$

The general picture that emerges from the our calculations is that at low pressures (< 20 GPa) NaZnF₃ in its perovskite and post-perovskite structure is the dominant compound in the Na-Zn-F system. Above that pressure NaF-rich (Na₂ZnF₄) and ZnF₂-rich (NaZn₂F₅) variants emerge as the most stable phases. Above 65 GPa they are predicted to decompose into binary fluorides. Inspection of the equation of states of the most stable phases in each of the three studied systems (Fig. S5 [40]) indicates that the driving force for these transitions is volume reduction. This is also connected with the increase of the coordination number of $Zn^{2+}$: the maximum value found for NaZnF₃ structure is 7, followed by 8 found for Na₂ZnF₄ and NaZn₂F₅, while for ZnF₂ in the PbCl₂ structure this value increases to 9. The coordination number of $Na^+$ (8) is the same in almost all of the compounds of the Na-Zn-F system.

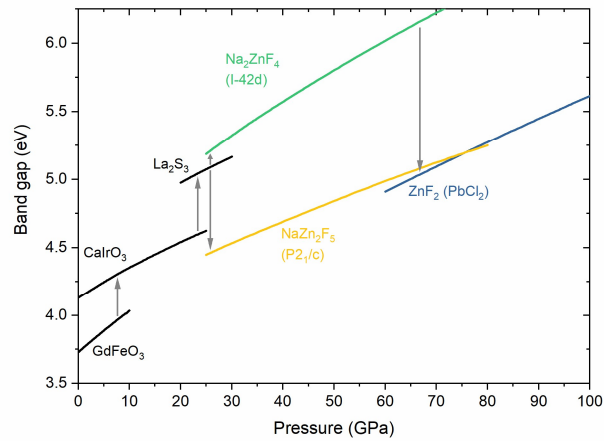

**FIG. 7** The pressure dependence of the band gap on NaZnF₃ phases, Na₂ZnF₄, NaZn₂F₅, and ZnF₂ (black, green, yellow, blue line, respectively. Arrows indicate band gap changes upon phase transitions and decomposition reactions.

Our calculations indicate that all of the members of the Na-Zn-F system are insulating at ambient and high pressure. For all Zn-bearing compounds the electronic band gap increases upon compression (Fig. 7). The rate of this increases is larger for NaZnF₃ phases and $I\bar{4}2d$ Na₂ZnF₄ (0.02 − 0.03 eV/GPa) compared to $ZnF_2$-rich stoichiometries, $P2_1/c$ $NaZn_2F_5$ and $PbCl_2$-type $ZnF_2$ (0.015 eV/GPa). While consecutive phase transition in $NaZnF_3$ and formation of $Na_2ZnF_4$ are connected with widening of the band gap, the opposite effect is seen for $NaZn_2F_5$ and $ZnF_2$ formation upon $NaZnF_3$ decomposition. We note that although the PBEsol functional underestimates the electronic band gap of $ZnF_2$, it models its pressure dependence similarly to more accurate meta-GGA functionals [52,68].

**TABLE I** Summary of the predicted pressures (in GPa) of the perovskite to post-perovskite and post-perovskite to post-post-perovskite phase transitions, as well as the decomposition of $ABX_3$ systems. The structure type of the post-post-perovskite phase ($La_2S_3$ or $Sb_2S_3$), as well as the type of decomposition products ($AX$, $BX_2$, $A_2BX_4$, $AB_2X_5$) are given in parenthesis. Data for $MgSiO_3$, $MgGeO_3$, $NaMgF_3$ comes from local density approximation (LDA) calculations [25], results for $NaZnF_3$ are from PBEsol calculations (this work).

| $ABX_3$ compound | $GdFeO_3 \rightarrow CaIrO_3$ | $CaIrO_3 \rightarrow La_2S_3/Sb_2S_3$ | Decomposition |
|---|---|---|---|
| $MgSiO_3$ | 80 | 1 300 ($La_2S_3$) | 750 ($A_2BX_4 + AB_2X_5$) |
| $MgGeO_3$ | – | 268 ($La_2S_3$) | 178 ($A_2BX_4 + BX_2$) |
| $NaMgF_3$ | 18 | 43 ($Sb_2S_3$) | 29 ($AX + AB_2X_5$) |
| $NaZnF_3$ | 8 | 23 ($La_2S_3$) | 26 ($A_2BX_4 + AB_2X_5$) |

As can be seen in Table I, $NaZnF_3$ undergoes the same phase transition sequence as $MgSiO_3$, but at much lower pressures. Moreover, it also undergoes the same compression-induced decomposition with the formation of $A_2BX_4$ in the $I\bar{4}2d$ structure and $AB_2X_5$ in the $P2_1/c$ structure. In terms of high-pressure behavior $NaZnF_3$ is more similar to $MgSiO_3$ than other previously proposed LPAs: $MgGeO_3$ and $NaMgF_3$. The former system exhibits the $CaIrO_3 \rightarrow La_2S_3$ phase transition at much higher pressure than $NaZnF_3$, and does not form $MgGe_2O_5$ upon decomposition. In $NaMgF_3$ the $CaIrO_3$ structure transforms to a $Sb_2S_3$ polymorph, and $Na_2MgF_4$ is not formed upon pressure-induced decomposition. The low pressures of phase transition/decomposition reactions predicted for $NaZnF_3$ might facilitate performing high-pressure and high-temperature experiments, which is important in the context of potential large activation barriers associated with the predicted transitions [69]. Exploration of the such barriers in the Na-Zn-F system, although of considerable interest, is beyond the scope of this study.

## IV. CONCLUSIONS

Our DFT calculations indicate that $NaZnF_3$ could serve as a good low-pressure analogue of $MgSiO_3$ exhibiting the same sequence of phase transitions, and pressure induced decomposition, but at pressures an order of magnitude lower. All of the structures and compositions stabilized by high pressure in the Na-Zn-F system are analogous to those predicted for Mg-Si-O. We predict that two novel compounds: $Na_2ZnF_4$ and $NaZn_2F_5$ can be obtained at ambient pressure ($Na_2ZnF_4$) or relatively low pressure of 19 GPa ($NaZn_2F_5$). We note that the latter compound is rare example of a $AM_2F_5$ (A

= alkali metal; M = M$^{2+}$) fluoride – such a composition is only found for M = Sn [70,71], Be [72], Pd [73,74], and Cu [75].

Given the relatively low pressures of the predicted phase transitions and decomposition reaction we hope for a fast experimental verification of the current results. We note, that Raman scattering experiments seem to be a good tool for exploring the high-pressure behavior of this system, as there are considerable differences in the Raman spectrum of Na-Zn-F phases (Fig. S6 [40]). Given that the ambient-pressure high-temperature approach was unsuccessful in the synthesis of Na$_2$ZnF$_4$ and NaZn$_2$F$_5$, other routes, such as mechanochemical synthesis by using high-energy ball milling (as applied in the case of Zn(BF$_4$)$_2$ synthesis) [76], or high-pressure high-temperature synthesis, with the use of a multi-anvil apparatus [77], might be pursued.

## ACKNOWLDEGEMENTS

D.K. acknowledges the support from the National Science Centre, Poland (NCN) within the SONATA BIS grant (no. UMO-2019/34/E/ST4/00445). This research was carried out with the support of the Interdisciplinary Centre for Mathematical and Computational Modelling at the University of Warsaw (ICM UW), under grant no. GA83-26. Z.M. gratefully acknowledges the financial support from the Slovenian Research Agency (research core funding No. P1–0045 Inorganic Chemistry and Technology).

Raman spectrum and structural parameters; comparison between $Na_2ZnF_4$, $Mg_2SiO_4$ and $CdMn_2O_4$.